\documentclass[article]{aa}  
\usepackage{natbib}
\bibpunct{(}{)}{;}{a}{}{,}
\usepackage{graphicx}
\usepackage{rotating}
\usepackage{txfonts}
\usepackage{multirow}

\def\h2{H$_2~$2.12$\mu$m}

\def\kms{km~s$^{-1}$}

\def\degrees{$\,^{\circ}$}
\begin{document}

\title{Spectro-astrometry  of V1515 Cyg \thanks{Based on service observations made with the WHT operated on the island of La Palma by the Isaac Newton Group in the Spanish Observatorio del Roque de los Muchachos of the Instituto de Astrofísica de Canarias.}} 

\author{  V. Agra-Amboage \inst{1}
          \and
          P. J. V. Garcia \inst{1}
}
\institute{Universidade do Porto, Faculdade de Engenharia, Departamento Engenharia Fisica, SIM Unidade FCT n\textordmasculine 4006, Rua Dr. Roberto Frias, s/n 4200-465, Porto, Portugal
            \email{vaa@fe.up.pt; pgarcia@fe.up.pt}
}
\offprints{V. Agra-Amboage}
\date{Received ; Accepted }

\abstract
{FU Orionis objects are a class of young stars with important bursts in luminosity and which show  evidence of accretion and ejection activity. It is generally accepted that they are surrounded by a Keplerian circumstellar disk and an infalling envelope. The outburst would occurs because of a sudden increase in the accretion rate.} 
{We aim at studying the regions closer to the central star in order to observe the signs of the accretion/ejection activity.} 
{We present optical observations of the H$_{\alpha}$ line using the Integral Field Spectrograph OASIS, at the William Herschel Telescope, combined with Adaptive Optics. Since this technique gives the spectral information for both spatial directions, we carried out a two-dimensional spectro-astrometric study of the signal. }
{We measured a clear spectro-astrometric signal in the North-South direction. The cross-correlation between the spectra showed a spatial distribution in velocity suggestive of scattering by a disk surrounding the star. This would be one of the few spatial inferences of a disk observed in a FU Orionis object. However, in order to fully understand the observed structure, higher angular and spectral resolution observations are required. V1515 Cyg appears now as an important object to be observed with a new generation of instruments to increase our knowledge about the disk and outflows structure in FU Orionis objects.} 
{}
\maketitle
%

\section{Introduction}

  FU Orionis objects (FUor) are a class of young stellar objects showing important outbursts, increasing in luminosity about 5 magnitudes and changing their spectral type in short timescales. The typical rise time is over one year, whereas the time scale for the lifetime of the high phase is decades. Two main theories were proposed to explain their nature. The first one favors a scenario where an unstable star is rotating near breakup, which would cause the outburst. A rapidly rotating G supergiant photosphere overlaid with a rising cooler shell could explain the observed spectral properties of these objects \citep{Herbig2003}. The second model, more widely accepted nowadays, considers a proto-stellar object surrounded by a Keplerian circumstellar disk and an infalling envelope. The outburst would occur because a sudden increase in the accretion rate through the disk. In this scenario, all young stars experience FUor phases during their evolution \citep{Hartmann1985,Hartmann1996}. Many studies during the last years investigated the accretion phenomenon in this class of objects from both an observational and theoretical point of view. Despite the effort the origin of the episodic accretion and outburst is poorly understood. A detailed review is presented by \citet{Audard2014}, including future promising directions.

Evidence of outflows in FUor objects have already been reported before by different authors and using different observational tracers, both in the optical and in the infrared \citep{Croswell1987}, showing P Cygni profile in different lines, such several hydrogen lines, lower excitation lines of neutral metals and TiO bands \citep{Hartmann1996} and continuum radio observations \citep{Rodriguez1992}. In particular for the Z\,CMa system (Herbig Be star and FUor object), \citet{Whelan2010} detect jets driven by each of the components in the [FeII] lines . At much smaller scales \citet{Benisty2010} detect collimated Br$\gamma$ from the Be component. Typical observed wind velocities are in the range of 300--400~\kms with mass-loss rates in the order of $\sim$10$^{-5}$--10$^{-6}$M$_{\sun}$~yr$^{-1}$, even if this is a very variable parameter from one source to another. In the case of V1515 Cyg, discovered by \citet{Herbig1977}, evidence of winds and outflows were found. Both \citet{Bastian1985} and \citet{Croswell1987} showed spectra with a clear H$_{\alpha}$ P Cygni profile and \citet{Croswell1987} determined a mass loss rate of 10$^{-5}$ M$_{\sun}$/yr from the spectral energy distribution. This value was then confirmed by \citet{Kenyon1991}. They compared the mid- and far-infrared SED to the predictions of standard accretion disk models and concluded that the presence of an infalling envelope in this object is needed to fit the model, but the presence of a cavity in the envelope, through which the central optical source is seen, is also required. More recently \citet{Green2006} obtained similar conclusions. Using IR Spitzer IRS observations, they also conclude that the envelope model requires an outflow hole with a large opening angle. They suggested that this cavity should be the result of the high mass-loss rate accompanying rapid accretion in the FUor outburst state.

Despite the evidence of the existence of winds/outflows and accretion activity  in these objects, a detailed study of their physical properties have been difficult to carry out with high resolution instruments manly due to the faintness of these objects and their distance to us, for instance V1515 Cyg is 1 kpc away \citep{Racine1968}. Here, we present a new approach through an Integral Field Spectroscopy study at medium spatial resolution which allows us to astrometricaly probe scales typically in the order of 100-1000 AU to the central star and carry out a spectroscopic study in the both spatial directions simultaneously.

 This article is organized as follows: in Sec.~\ref{sec:obs} we describe the observations and the data reduction, in Sec.~\ref{sec:results} we present our results in terms of detection and reality of the detection. Section~\ref{sec:discussion} we propose some explanations for the observed signal in the context of the previous works and summarize our conclusions.

\section{Observations and data reduction}
\label{sec:obs}

 Observations of V1515 Cyg were obtained at the William Herschel Telescope with the Integral Field Spectrograph OASIS combined with the Adaptive Optic module NAOMI mounted at the Nasmyth focus. The summary of the observing run is presented in Table~\ref{tab:obs}. Two different configurations were used, HR638 and HR667, with a final spatial sampling of 0\farcs19, which gives a FOV of 9\farcs5$\times$7\farcs2. The first configuration, HR638, provides a spectral resolution of 3190 in a wavelength range between 6210\AA~and 6550\AA, sampled at 0.96\AA/px. The second configuration, HR667,  provides a spectral resolution of 3340 for a wavelength range of 6490--6840\AA, sampled at 0.97\AA/px. For each configuration two exposures of 1800s each were taken. After AO correction, the effective spatial resolution achieved is 0\farcs7 (Gaussian core FWHM).

 Data reduction was carried out using the XOasis software provided by the OASIS team in Lyon (France)\footnote{http://www.cfht.hawaii.edu/Instruments/Spectroscopy/OASIS/Reduc/}. The raw images were bias subtracted. In order to extract the spectra, the positions of the micro-lenses on the CCD were obtained by a micro-pupil image taken during the observation run. This images consists in illuminating the instrument with a source of uniform light with the dispersive element removed.
In this image a Gaussian fit is carried out to obtain the center of each micro-pupil. Then, a cross-dispersion profile is evaluated for each spectrum and then fitted using an algorithm which computes an optical model of the instrument dispersion on the detector. The positions of the spectra ridges are obtained from these cross-dispersion profiles. 
For every lens and every wavelength, the CCD  pixels are summed up over a total width of five pixels centered on the positions determined before. Each summed pixel has a weight which is computed with an optimal algorithm  \citep{Horne1986},
using the photon noise of the fraction of the signal on the pixel and the known CCD readout noise.
Finally an extraction mask is created and the extracted spectra are arranged in a data cube with the two spatial dimensions and a first estimation of the wavelength vector. Using a Neon lamp a refined wavelength calibration is carried out by fitting the residuals of the first estimation, made in the previous step, to a second-order polynomial. The low frequency both spatial and spectral flat effects are corrected in the same step using a sky flat and the continuum frames (images where the instrument is illuminates with an uniform light).
 Then, the data cube is corrected from cosmic rays. 
For each spectrum, the integrated flux of the spectrum  and of the neighbors inside a region, where we can assume spectral similarity, usually twice the size of the spatial sampling, are computed. The neighbors spectra are normalized to the central one and the median of the spectra is calculated. The difference between the central spectrum and the median spectrum is computed (median-filtered difference) and compared with the un-filtered difference. A sigma-clipping is then performed on the resulting spectrum. The pixels above 30 times sigma are replaced, in the central spectrum, by the median value.
The flux calibration was carried out using the standard star image, over which all the previous steps were also carried out. 
Finally, the data cubes are  re-sampled to a square spatial sampling grid.

\begin{table}
\caption{Summary of observations taken on the 3$^{rd}$ September 2005. For both configurations a sampling of 0\farcs26 was used. The configurations HR667 was used to observing the H$_{\alpha}$ line and HR638 for the [OI].}\label{tab:obs}
\begin{tabular}{ccccc}
\hline
\hline
Object & Configuration & UT & T$_{exp}$ (s) & Airmass\\
\hline
V1515 Cyg & HR638 & 01:30 & 1800 & 1.28\\
V1515 Cyg & HR638 & 02:01 & 1800 & 1.39\\
Neon Lamp & HR638 & 02:34 & 12    &    \\ 
Standard star & \multirow{2}{*}{HR638} & \multirow{2}{*}{04:51} & \multirow{2}{*}{1500} & \multirow{2}{*}{1.23}\\
(G191-B2B)    &       &      &        &    \\
Neon Lamp & HR638 & 04:48 & 12    &    \\ 
\hline
V1515 Cyg & HR667 & 00:24 & 1800 & 1.12\\
V1515 Cyg & HR667 & 00:55 & 1800 & 1.18\\
Neon Lamp & HR667 & 01:26 & 3    &     \\ 
Standard star & \multirow{2}{*}{HR667} & \multirow{2}{*}{05:17} & \multirow{2}{*}{1500} & \multirow{2}{*}{1.18}\\
(G191-B2B)    &       &      &        &    \\
Neon Lamp & HR667 & 05:43 & 12   &     \\ 
\hline
\end{tabular}
\end{table}

\section{Results}
\label{sec:results}

 \subsection{Spectro-astrometry}
 \label{subsec:detection}

\begin{figure}
\centering
\includegraphics[width=0.4\textwidth]{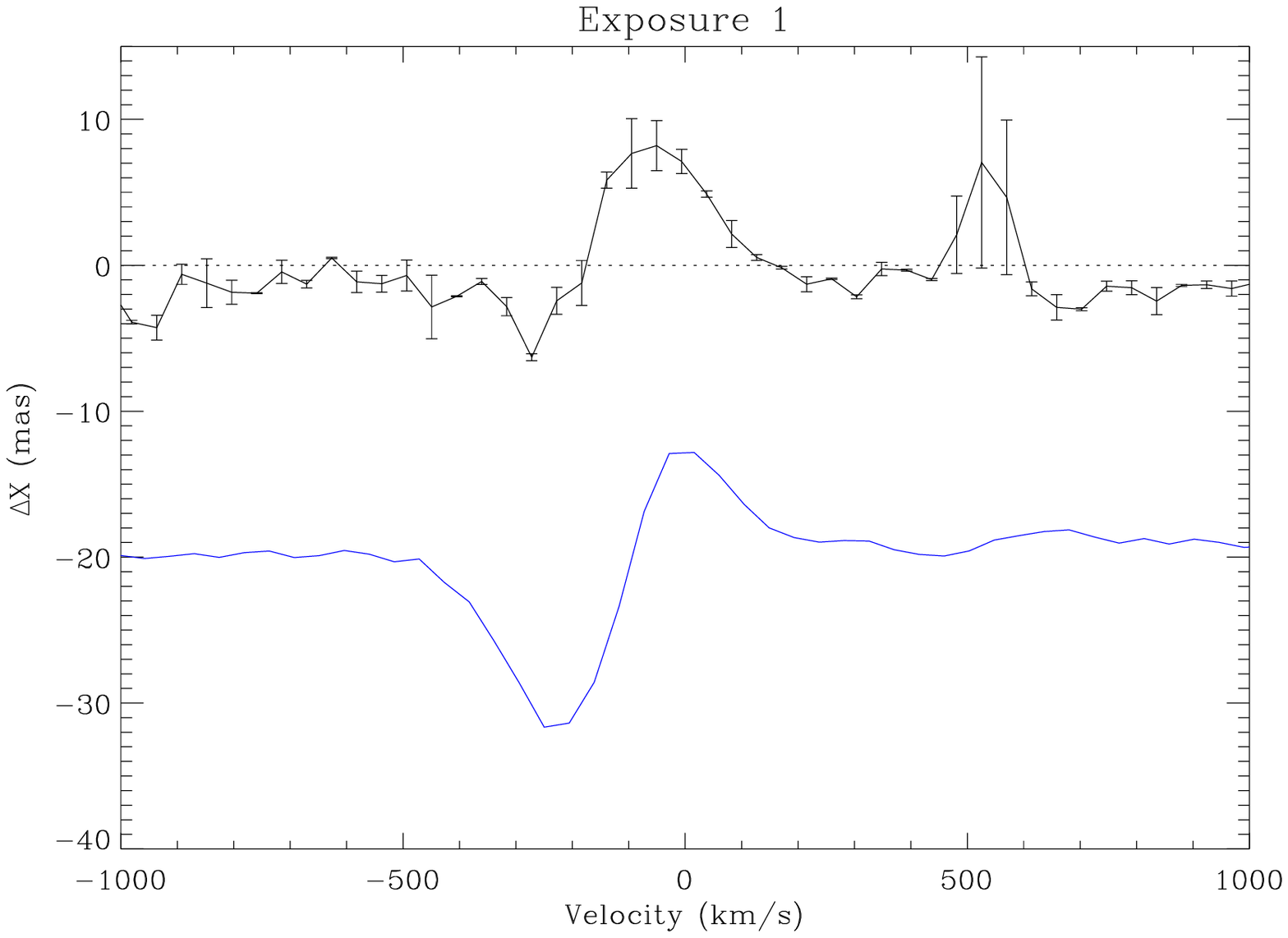}
\includegraphics[width=0.4\textwidth]{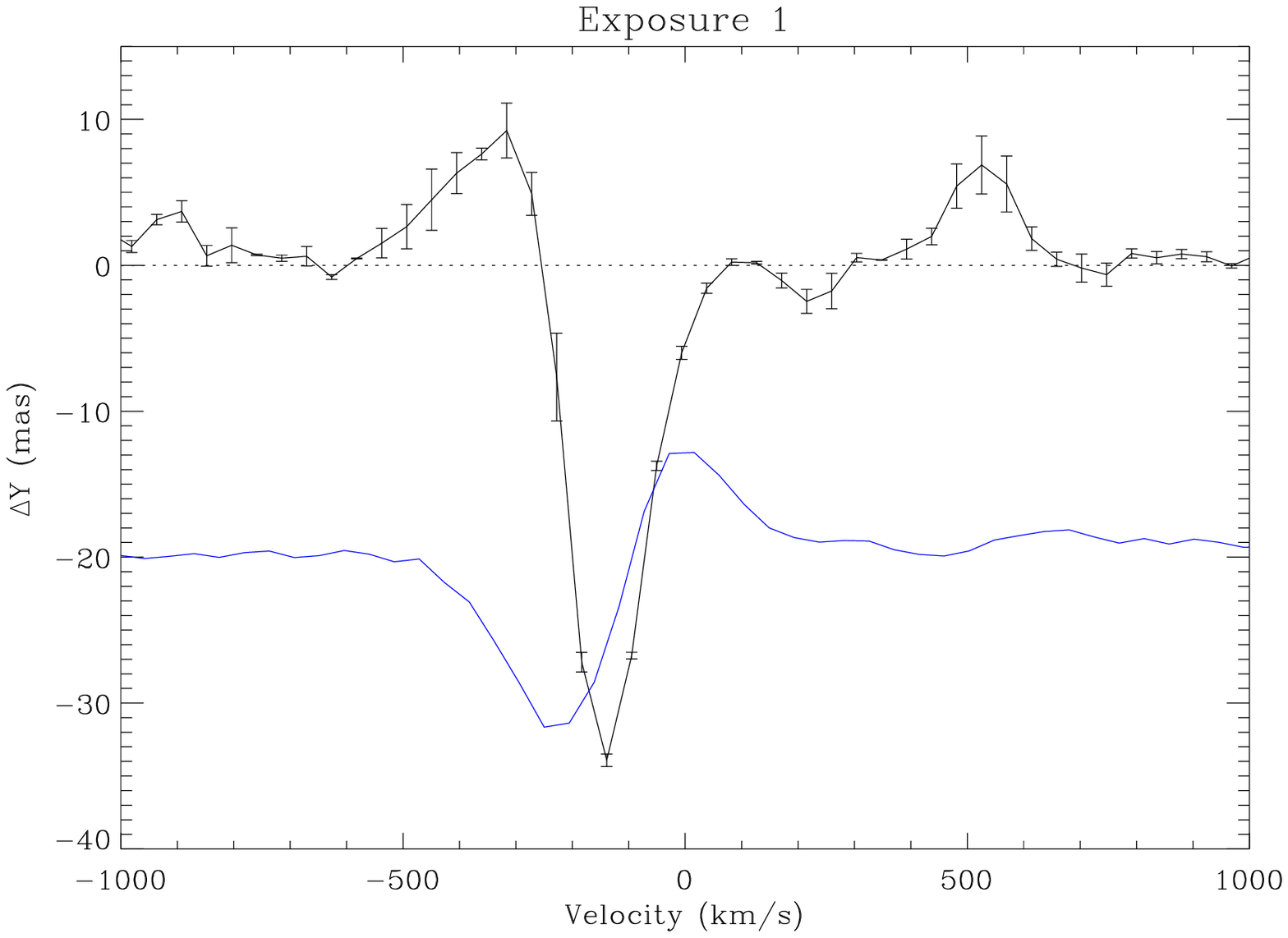}
\includegraphics[width=0.4\textwidth]{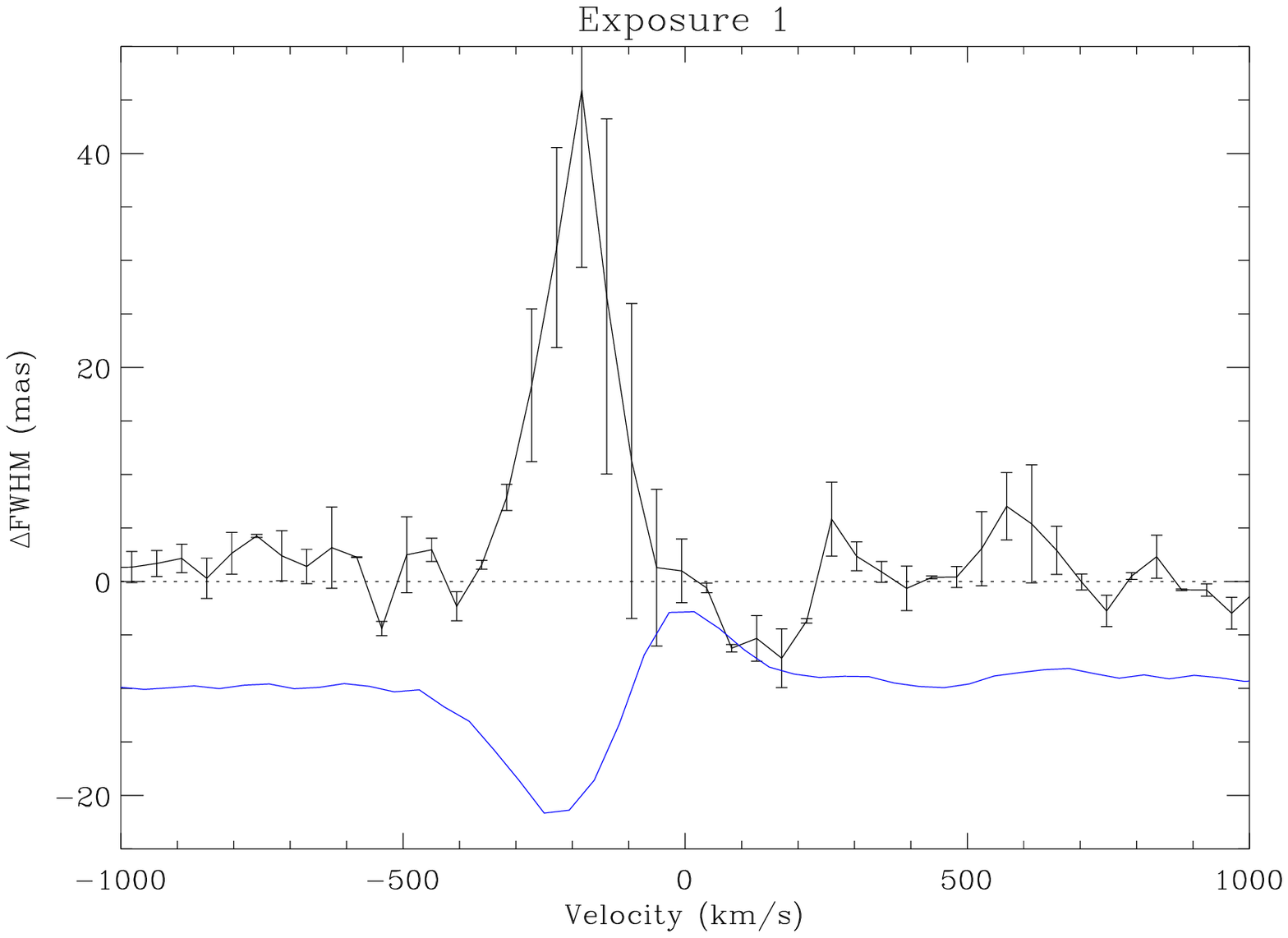}
\caption{For the first individual exposure, gaussian centers of the spectro-astrometric signal as a function of the velocity in both the W-E (East for $\Delta X > 0$) and S-N (North for $\Delta Y >0$) directions, in top and middle panel respectively. Bottom panel shows the differential FWHM, see text. The H$_{\alpha}$ line profile is over-plotted in blue showing a clear P Cygni profile.}
\label{fig:spectroastrometry69}
\end{figure}

\begin{figure}
\centering
\includegraphics[width=0.4\textwidth]{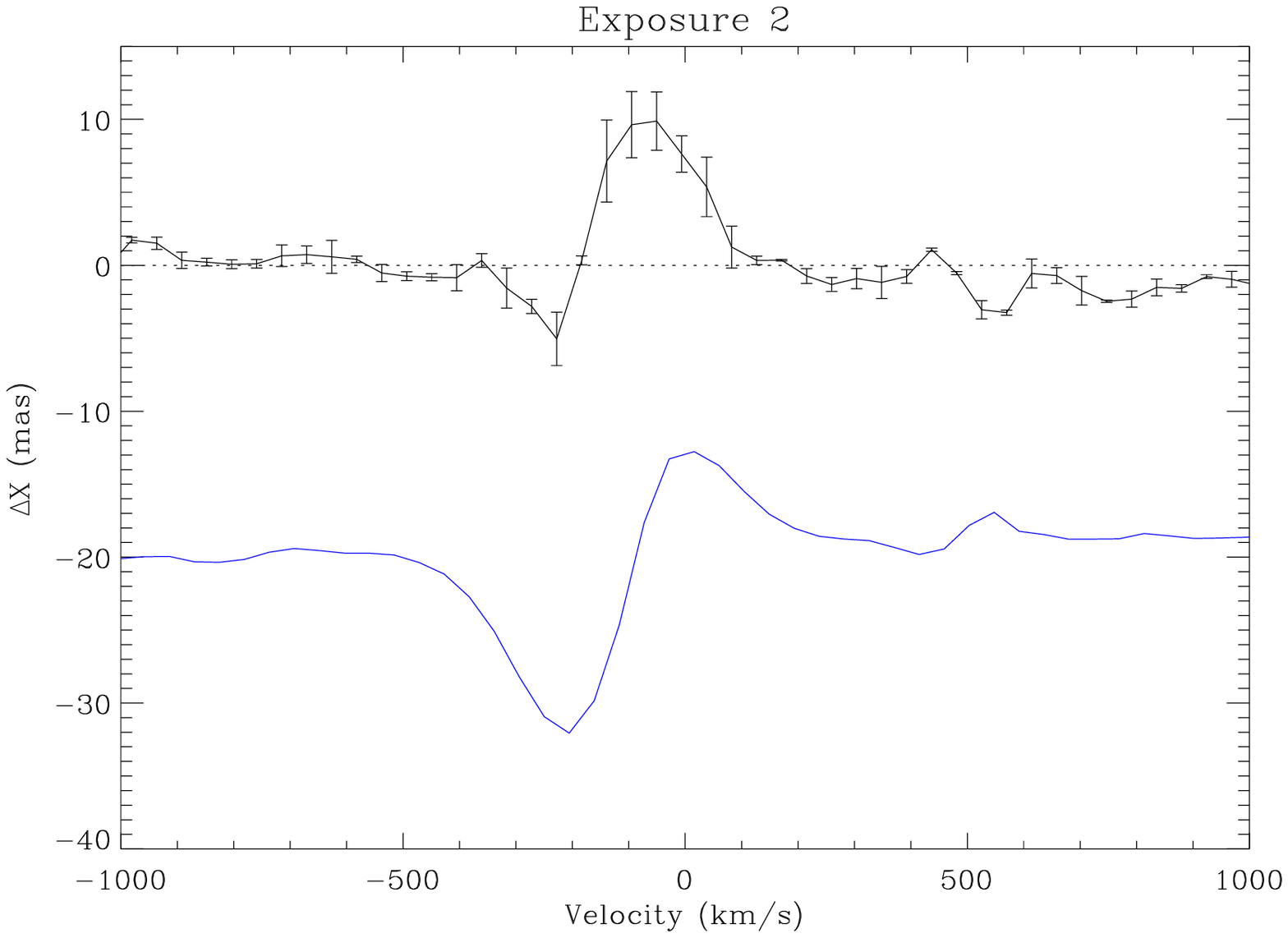}
\includegraphics[width=0.4\textwidth]{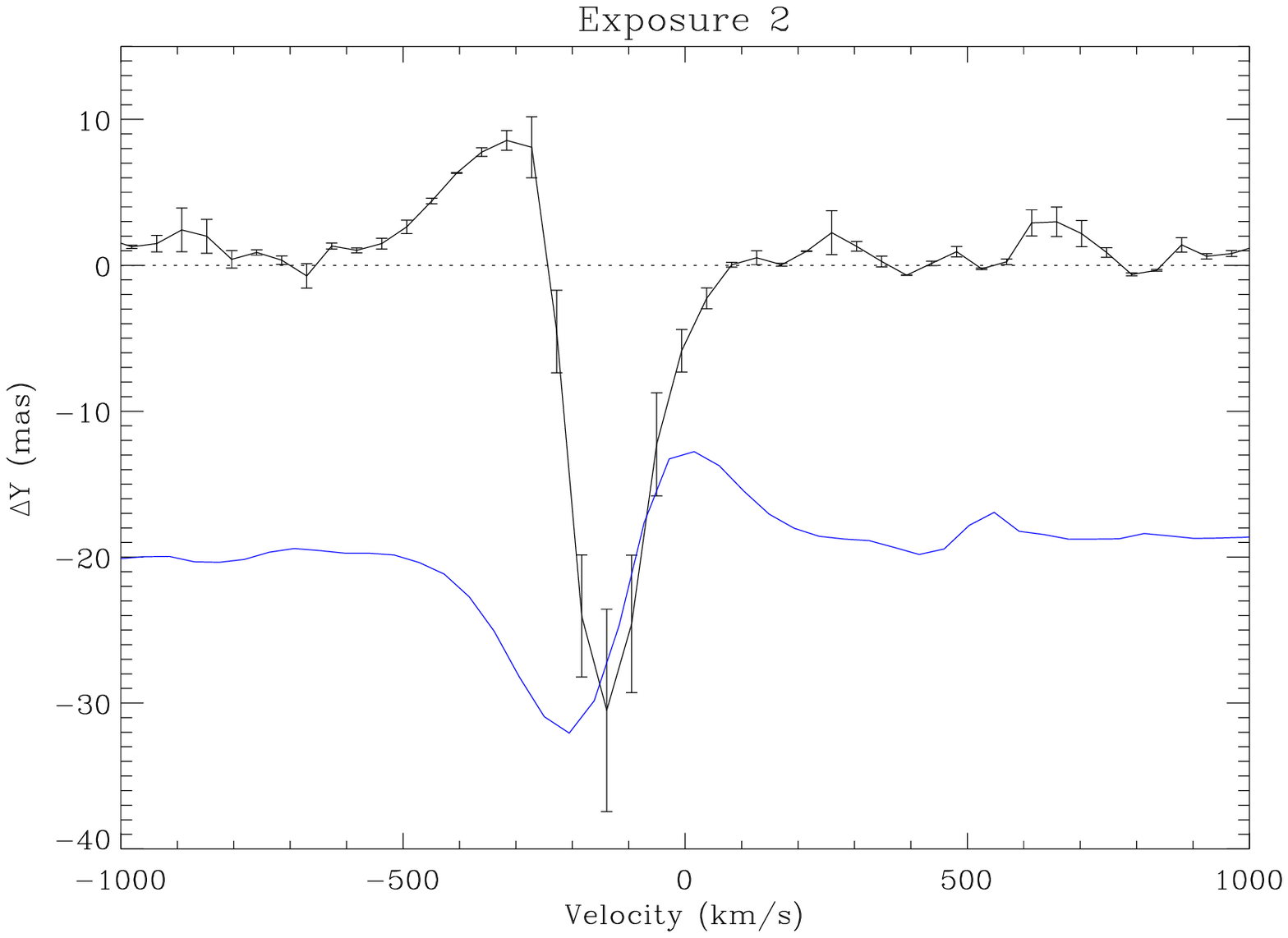}
\includegraphics[width=0.4\textwidth]{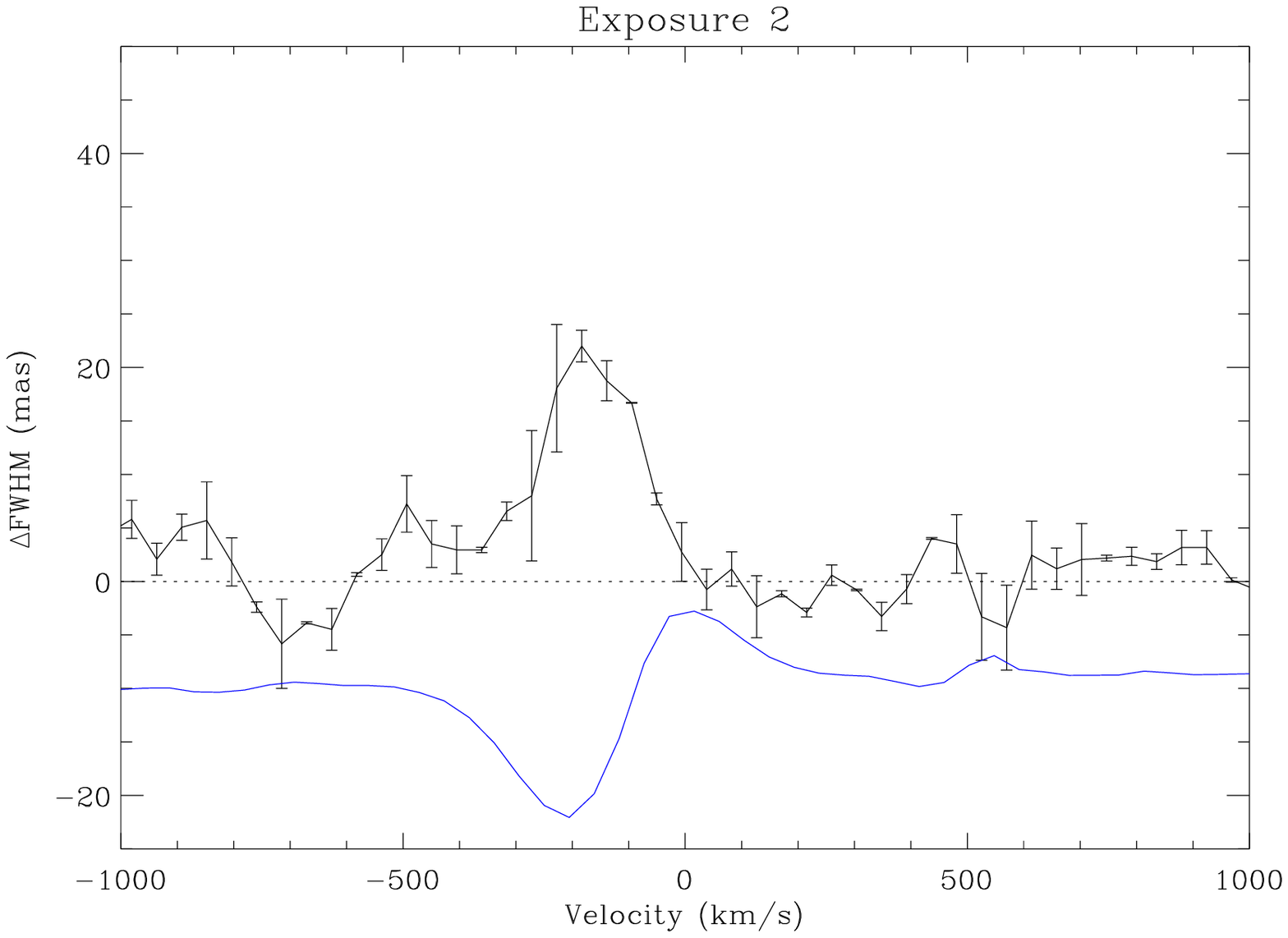}
\caption{Same as Fig.~\ref{fig:spectroastrometry69} for the the second individual exposure taken.}
\label{fig:spectroastrometry70}
\end{figure}

 The spectro-astrometry technique measures the astrometric position by a Gaussian fit of the spatial profiles for each individual wavelength \citep[e.g.][]{Whelan2008}. In the case of integral field spectroscopy, where the spectral information is given for each 2D spatial point, this technique can be applied to both spatial directions \citep[see as example][]{Davies2010}. Here, a 2D circular gaussian is fitted to each image of the data cube, with a Levenberg-Marquardt algorithm. The wavelength dependent gaussian centres and gaussian FWHM are the spectro-astrometric signal. Only the differential signals are presented, a linear contribution was removed by fitting a straight line to the data. Fig.~\ref{fig:spectroastrometry69} and Fig.~\ref{fig:spectroastrometry70} show, for each individual exposure, the detected spectro-astrometric signal in the spatial directions for the spectral region around the  H$_{\alpha}$6562\AA~line. Top panels show the horizontal direction on the detector, oriented West-East (positive $\Delta X$ is East), the middle ones refer to the vertical direction, oriented from south to North, and the bottom panels show the differential FWHM from the fit. The profile of the emission, arbitrarily normalized in order to include it in the plot, is also plotted in blue.  The wavelength scale is converted to a radial velocity scale taking the vacuum wavelength of  H$_{\alpha}$ as the reference velocity and a heliocentric correction for V1515Cyg of $-12 \pm2$ km/s \citep{Herbig1977}.

The P Cygni profile previously observed in this object \citep{Bastian1985} is clear in our data too. A small spectro-astrometric signal is detected in the horizontal direction. However, in the vertical direction, considering the error bars, the signal is clear at blueshifted velocities of $\sim -100$ km/s. For the same velocities, in the absorption part of the PCygni profile, the FWHM increases. The differential $\Delta$FWHM signals for the two exposures seem to have a different amplitude. However, the errors in the first exposure are high and this is a $\lesssim 2\sigma $ effect.

 We also looked for detection in the [OI]$\lambda$6300\AA\ line, but no [OI] emission is present in the spectrum. We derive an upper limit of detection of 21.4$\times$10$^{-19}$ Wm$^{-2}$\AA$^{-1}$arcsec$^{-2}$, estimated as the 3-sigma value in the spectrum placed at the continuum center. A spectro-astrometric study revelead no signature, as expected.


\subsection{Test of the spectro-astrometric signal}\label{sec:test}
 
 The slit effect does not affect OASIS-like integral field spectrograph \citep{Bacon1995}. We thus analyzed the two main bias which can cause the observed spectro-astrometric signal: 1) flux contamination from the neighboring lenses during the extraction of the spectra and 2) a deficient wavelength calibration. 

\paragraph{{\it 1) Flux contamination from neigboring lenses.}} As explained in Section~\ref{sec:obs}, by default each spectra is extracted by summing in a five pixels aperture. In the case of a contamination from neighboring lenses, the effect would be minimized and even eliminated by choosing a smaller aperture. However summing in a three pixel aperture and in a one pixel aperture, the observed effect remained. Hence, we ruled out this as cause of the spectro-astrometric signal.

\begin{figure}
\centering
\includegraphics[width=0.5\textwidth]{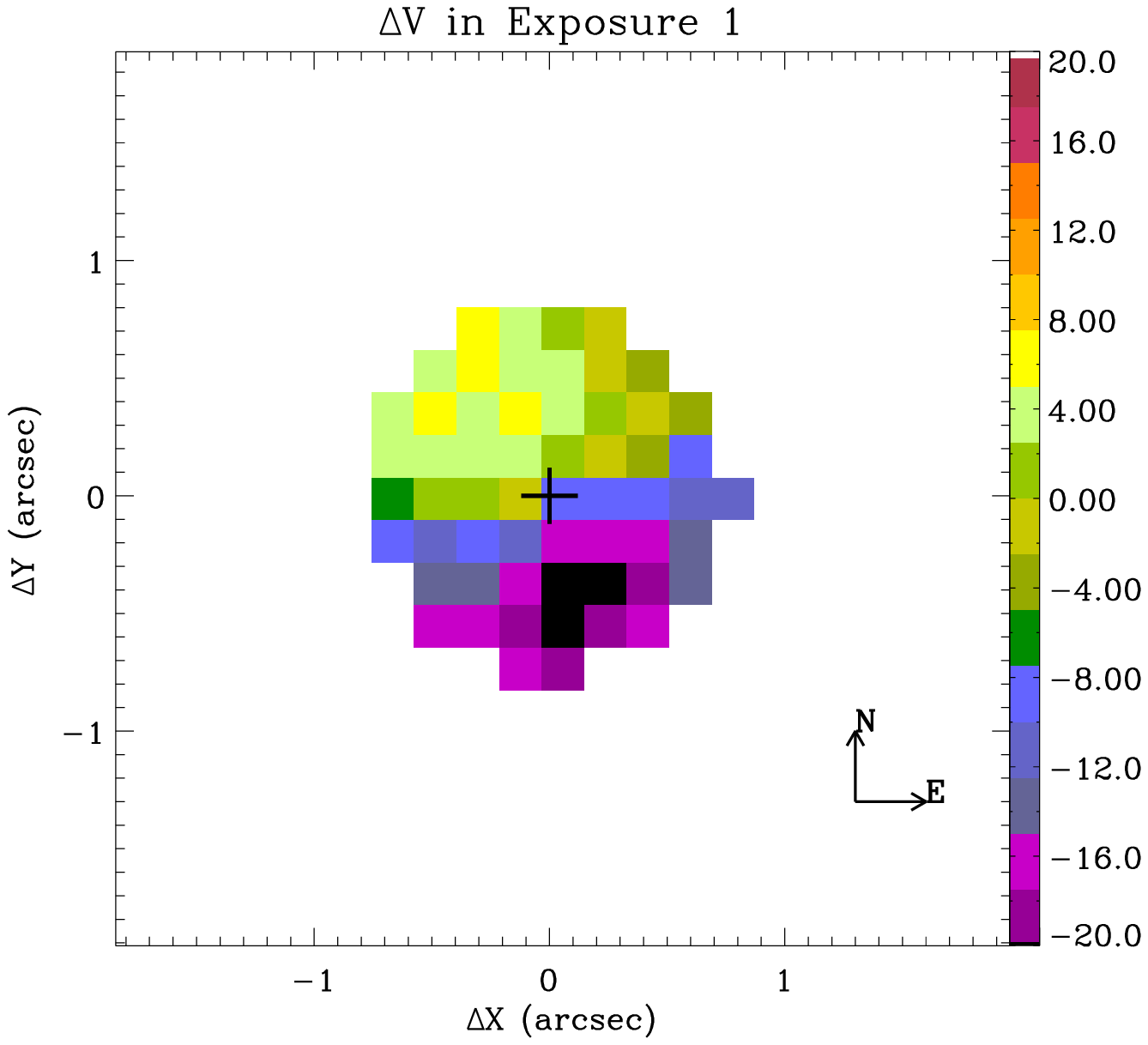}
\includegraphics[width=0.5\textwidth]{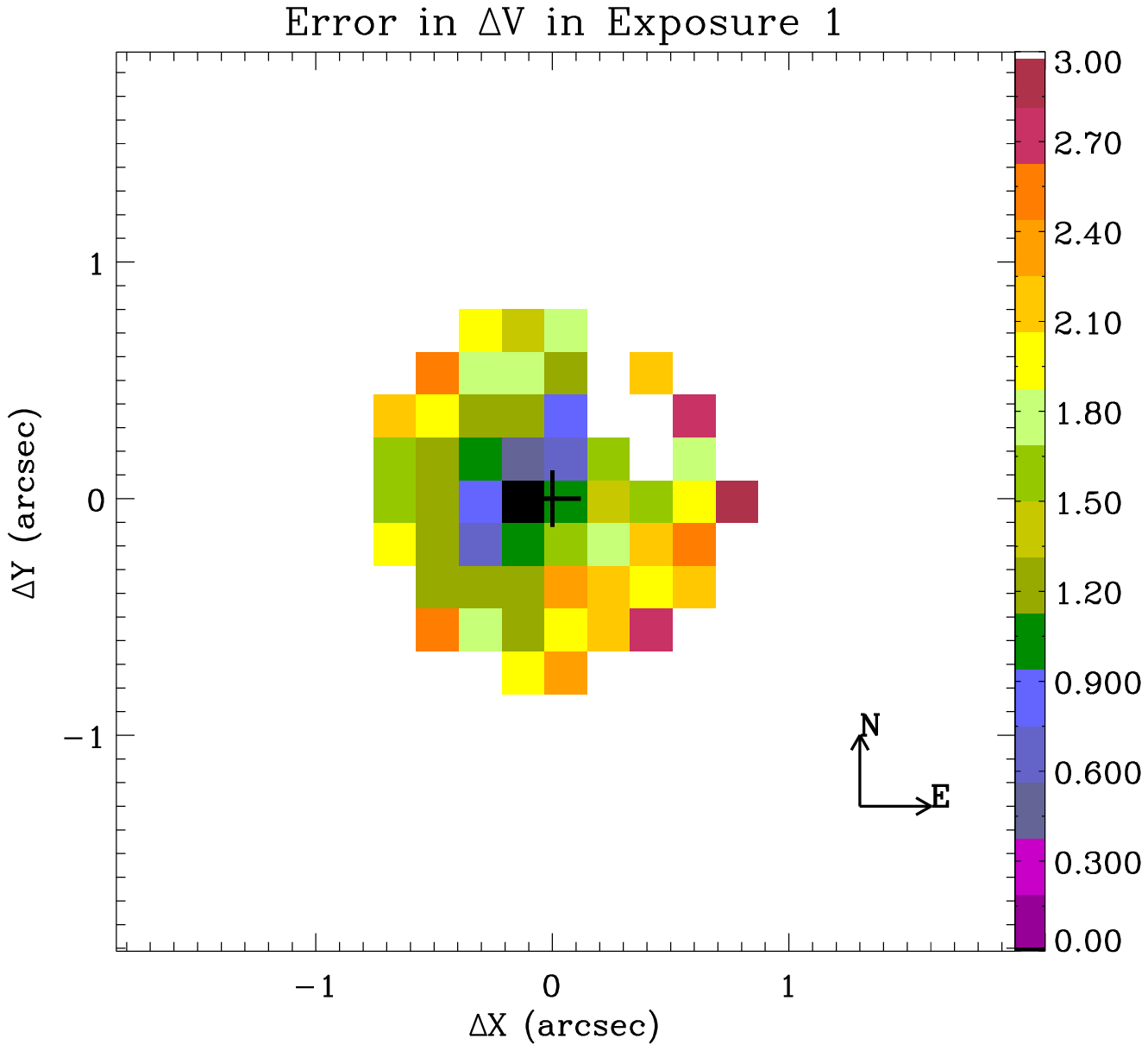}
\caption{Top: Shift in velocity obtained from a cross-correlation of the P Cygni profile between the spectrum at the position of the star and each individual spectra in the field, for exposure 1. Color scale is \kms. The continuum emission center is marked as a black cross. The physical orientation is also given. Bottom: Error in the shift also in \kms. }
\label{fig:cross-correlation1}
\end{figure}

\begin{figure}
\centering
\includegraphics[width=0.5\textwidth]{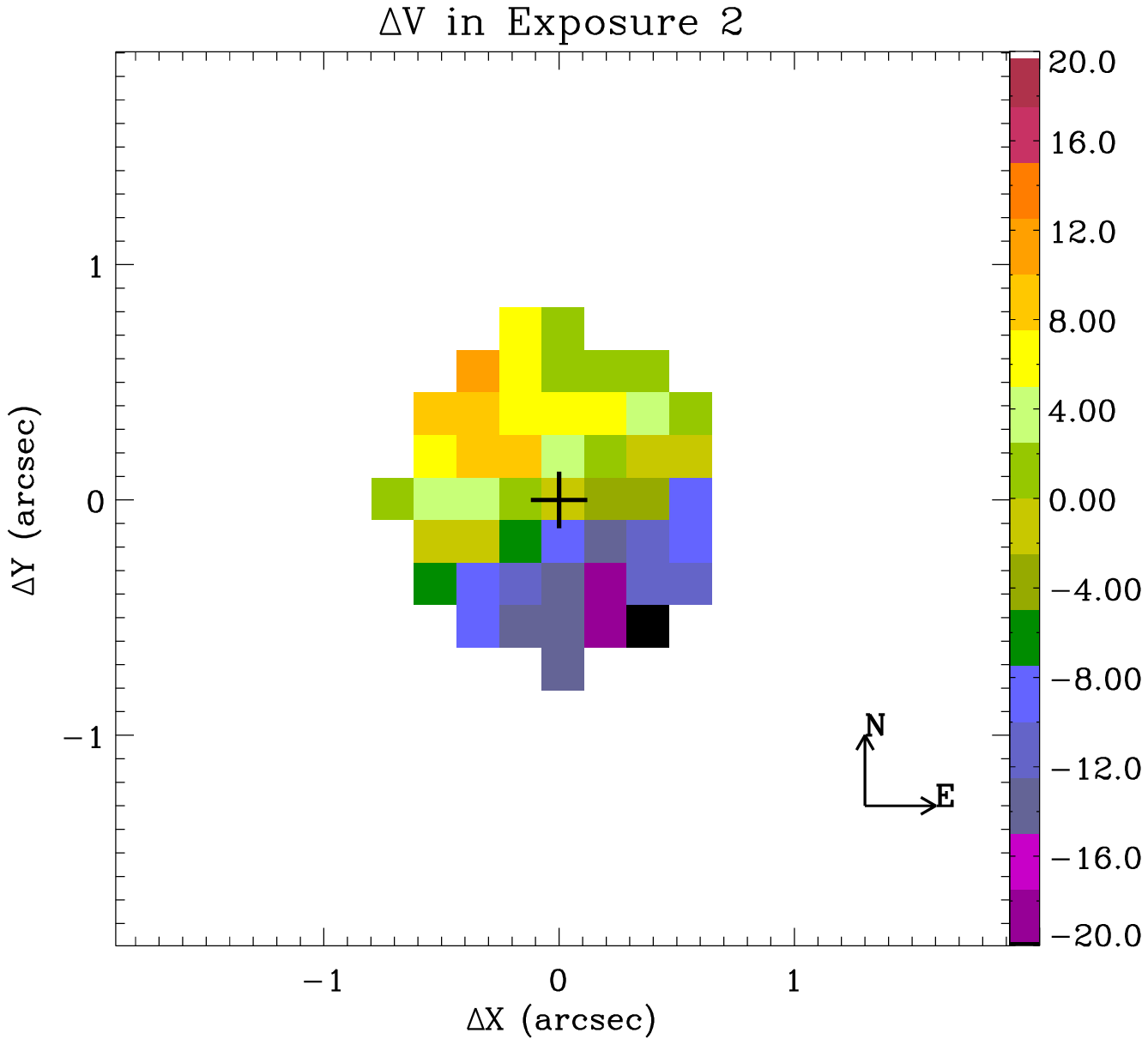}
\includegraphics[width=0.5\textwidth]{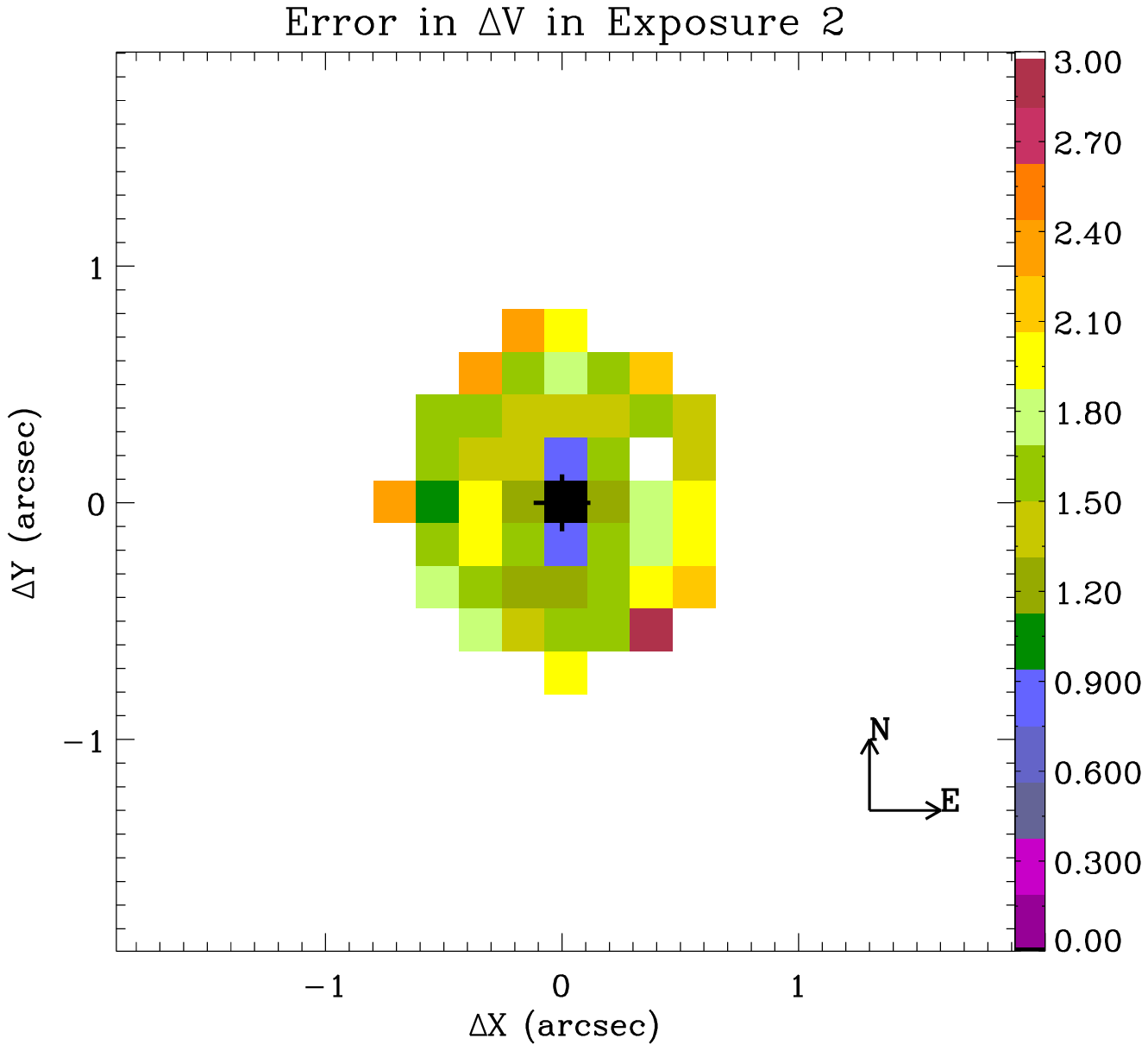}
\caption{Same as Fig.~\ref{fig:cross-correlation1} for exposure 2.}
\label{fig:cross-correlation2}
\end{figure}

\paragraph{{\it 2) Deficient wavelenght calibration.}} Model fitting between each spectrum in the field and the one at the stellar position was carried out. The spectra in each lens was modelled as a velocity shifted and scaled version of the brightest lens spectra. This two parameter model was fitted with a Levenberg-Marquardt algorithm. Only the points around the H$\alpha$ profile (with $|v|<10^3~\mathrm{km~s}^{-1}$) were considered in the fit. Top panels in Figs~\ref{fig:cross-correlation1} and ~\ref{fig:cross-correlation2} show the spatially distribution of the measured shift for each of the exposures taken. The continuum emission center corresponds with the position of the spectrum used as reference. A clear structure is seen in the center of the field with an approximate size of $\sim$2\arcsec~in both spatial directions. We identify two clear distinct regions, one showing redshifted velocities up to $\sim$12 \kms~in exposure 2, and the other blueshifted, almost symmetrical to the previous one, with values typically between -16 and -8 \kms. Bottom panels show the estimated errors for the velocities shift at each point. A quite well defined imaginary line of PA$\sim$53\degrees separates the two regions of opposite velocities.  An alternative check was made with a cross-correlation routine, CRSCOR  in IDL, and the same results were found.

\begin{figure}
\centering
\includegraphics[width=0.5\textwidth]{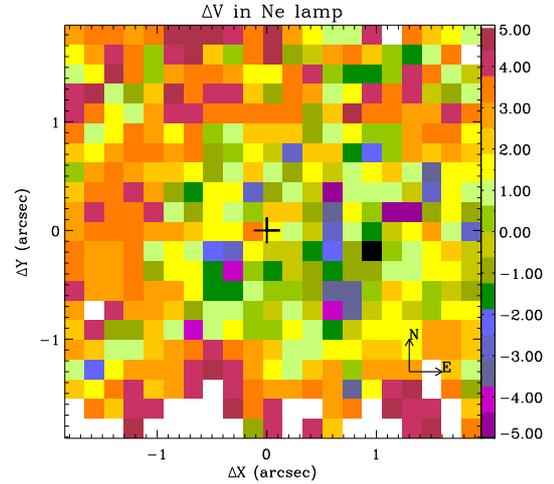}
\caption{Same as  Fig.~\ref{fig:cross-correlation1} but for the Neon arc  used to wavelength calibration.}
\label{fig:cross-correlationcalib}
\end{figure}

In the case of a problem in the wavelength calibration process, since the same procedure was used for all the cubes of the data set, we expect to find a similar structure in other images of the run. Hence we applied the same study to the image of the Neon arc used to carry out the wavelength calibration, Fig.~\ref{fig:cross-correlationcalib}. The arcs included five emission lines, which were fitted with Moffat functions with free amplitude and centre, but with fixed shape (i.e. $R$ and $\beta$ \citep{Moffat1969}). The wavelength calibration correction for each lens was computed with a 2nd order polynomial. The velocity corrections for each lens, at H$\alpha$, were compared with the brightest lens and no effect found. These results shows that the effect is not due to a wavelength calibration bias. 



\section{Discussion and conclusion}
\label{sec:discussion}

  We showed a clear spectro-astrometric signal observed in the H$_{\alpha}$ emission of the FUor star V1515 Cyg. The signal was tested for bias and we conclude that the observed velocity structure observed in H$_{\alpha}$, shown in Figs.~\ref{fig:cross-correlation1} and~\ref{fig:cross-correlation2}, is real, with a physical origin and characteristic of the H$_{\alpha}$ line, since none other emission is detected in other lines typically tracing outflows such as [OI]$\lambda$6300\AA, [NII]$\lambda$6584\AA~or [SII]$\lambda$$\lambda$6716,6716. 
The velocity distribution of the observed structure suggests a rotating disk around the young star, that we can observe because of the scattering of the light by the disk. The FUor objects are assumed to be similar to  T Tauri stars, consisting in a surrounding disk, maybe a remaining envelope, and an outflow. In fact what we observe here is very similar to the observed velocity gradients in $^{12}$CO images of protoplanetary disks with keplerian rotation in several T Tauri stars \citep[e.g.,][]{Simon2000}. In addition, several previous works on V1515 Cyg concluded about the need of an infalling envelope in order to fit the observed SED, in fitting the excess at 10 $\mu$m, and to explain the low K-band visibilities observed \citep{Millan-Gabet2006,Green2006,Zhu2008}. The models by \citet{Zhu2008} and by \citet{Green2006} also require the existence of an outflow cavity in the envelope. As shown by \citet{Kospal2011} this star is also surrounded by a nebula. Hence, the environmental medium in this star is a plausible option to produce scattering by dust either on the disk or on the envelope. 

 However, where is the light coming from? It may be stellar radiation scattered in a disk in keplerian rotation, scattering by a dusty halo or it may be the light coming from an outflow. We briefly discuss these options in base of the results presented here.

\paragraph{{\it 1) Disk in keplerian rotation:}} as said before, the observed morphology of the structure strongly suggest scattering in a disk in keplerian rotation. However, for the 1 kpc distance of this object and a typical velocity ($V_k$) of 10 \kms~observed in Figs.~\ref{fig:cross-correlation1} and~\ref{fig:cross-correlation2} at a distance for the star of 1\arcsec ($r$), the expected central star mass, $M_*=V_k^2\times r/G$, is in the order of $\sim$100 M$_{\sun}$ which is too large for this class of objects. Even in the regions where the observed velocity is $\sim$4 \kms, the derived star mass is $\sim$20 M$_{\sun}$ which is still too high. In the case of FU Ori a mass of 0.3 M$_{\sun}$ has been estimated by \citet{Zhu2007}. Assuming a higher mass of $\sim$2 M$_{\sun}$, in the order of the highest masses observed in T Tauri stars \citep{Calvet2004}, the keplerian velocity only explains $\sim$1 \kms~of the observed values. An additional contribution is needed.

 \paragraph{ {\it 2)  Dusty halo:}} The scattering light can also be produced by dusty haloes, remnants of the original envelopes.  \citet{Wheelwright2010} 
carried out a spectro-astrometric study over several Herbig Ae/Be (HAe/Be) stars. They observed in several of them a behavior similar of that observed here in V1515 Cyg. Large FWHM features occurred over absorption features in the emission profiles within small positional signatures (bottom and top panels of Figs.~\ref{fig:spectroastrometry69} and~\ref{fig:spectroastrometry70}). They suggested that these signatures could trace extended structures which scatter the light, such as a disk/stellar wind \citep{Azevedo2007}, haloes  \citep{Leinert2001,Monnier2006}, or nebulosity. They considered the presence of dusty haloes and carried out a simple model. In spite of the limits of their model, such as for example the unknown amount of light scattered and the extend of the halo, they reproduced quite well the spectro-astrometric signatures observed for the PCygni profile observed over the H$_{\alpha}$ line in AB Aur (their Fig.~4). In our case, however, the positional signature in the vertical direction is very different from the horizontal one and comparable with the FWHM features. This asymmetry could be caused by an outflow or by projection effects. 

\paragraph{ {\it 3)  Outflow:}}  \citet{Azevedo2007} showed that the observed spectro-astrometric signal could also result from a wind. Some authors favor a nearly pole-on orientation of the system \citep{Kenyon1991,Millan-Gabet2006,Zhu2008}, in which case we would be seeing the outflow coming to us through the predicted cavity. The peak at $\sim$8-10 \kms  (Figs.~\ref{fig:cross-correlation1} and~\ref{fig:cross-correlation2}) could be caused by a shock in the outflow itself or against the envelope cavity on the disk or the ambient medium. To explain the rest of the velocity structure, one simple explanation is that we are seeing 
the projected $\Delta V$ produced by rotation of an inner disk-wind. Disk-wind models predict rotation of the outflow, points at different sides of the jet with different velocities would be scattered by the disk or an envelope producing an observed $\Delta V$. 
The projected $\Delta V$ of rotation from the outflow will be $\sim$10 \kms which corresponds to the lower typical velocity differences measured by \citep{Coffey2007} in the case of DG Tau and interpreted as possible rotation signatures. However, for the regions with lower observed velocities the derived velocities are too low for relating them with rotation signatures.\\

  Unfortunately, this object is very faint, the signal to noise of our H$_{\alpha}$ line and the strong stellar continuum emission, do not allow a more detailed analysis of this detection in order to extract more quantitative results and the disk parameters. In addition, the exact inclination of the system is unknown, so projection effects could play an important role in the interpretation.

 In conclusion, the spatial distribution of the velocity structure observed in H$_{\alpha}$ suggests scattering by a disk/envelope surrounding the star. In that case, this would be one of the few spatial inferences of disks in FUor. Unfortunately our resolution does not allow a detailed parametrization of the emission observed in order to satisfactorily explain the velocity values  observed. Observations at higher angular and spectral resolution are required. Since little is known about the exact structure of the disk and outflows in this class of objects, V1515 Cyg appears now as an excellent candidate for future instruments with higher angular and spectral resolution.


\begin{acknowledgements}
    This research was partially supported by FCT-Portugal through Project PTDC/CTE-AST/116561/2010. Vanessa Agra-Amboage also acknowledges financial and travel support through the Funda\c{c}\~{a}o para a Ci\^{e}ncia e Tecnologia (FCT) under the contract SFRH/BPD/69670/2010. We thank Dr. Sylvie Cabrit and Dr. Catherine Dougados for their comments on the paper.
\end{acknowledgements}

\bibliographystyle{aa}
\bibliography{biblio}

\end{document}